\documentstyle[twocolumn,pra,aps,psfig]{revtex}

\begin{document}
\title{Frequency up-converted radiation from a cavity moving in vacuum}
\author{A. Lambrecht$^a$, M.T. Jaekel$^b$ and S. Reynaud$^a$}
\address{$^{a}$Laboratoire Kastler Brossel\thanks{%
Laboratoire de l'Ecole Normale Sup\'erieure et de l'Univer\-sit\'e Pierre et 
Marie Curie associ\'e au CNRS}, Universit\'e Pierre et Marie
Curie, case 74, 4 place Jussieu, 75252 Paris, France\\
$^b$ Laboratoire de Physique Th\'eorique de l'Ecole Normale Sup\'erieure \thanks{%
Laboratoire du CNRS associ\'e \`a l'Ecole Normale Sup\'erieure et 
\`a l'Universit\'e Paris Sud}, 24 rue Lhomond, 75231 Paris, France}
\date{to appear in {\sc European Physical Journal} {\bf D3}, 1998}
\maketitle

\begin{abstract}
We calculate the photon emission of a high finesse cavity moving in vacuum.
The cavity is treated as an open system. The field initially in the vacuum
state accumulates a dephasing depending on the mirrors motion when bouncing
back and forth inside the cavity. The dephasing is not linearized in our
calculation, so that qualitatively new effects like pulse shaping in the
time domain and frequency up-conversion in the spectrum are found.
Furthermore we predict the existence of a threshold above which the system
should show self-sustained oscillations.

{\bf PACS: } 42.50.Lc - 03.70.+k - 12.20.-m
\end{abstract}

Vacuum field fluctuations exert radiation pressure on scatterers in vacuum.
For a pair of mirrors at rest this effect is well known as Casimir effect 
\cite{Casimir48}. When a mirror is moving radiation pressure of vacuum
fluctuations leads to a dissipative force which opposes itself to the
mirrors motion. This force is known to arise as soon as the mirror has a
non-uniform acceleration \cite{FullingDavies76}. Accordingly the
electromagnetic field remains not in the vacuum state but photons are
emitted by the scatterer into vacuum \cite{Moore70}. Radiation from a moving
mirror and the associated radiation reaction force imply that dissipative
effects are associated with the motion of mirrors in vacuum, although this
motion has no further reference than vacuum itself. Since these effects
challenge the principle of relativity of motion in vacuum, it would be very
important to obtain experimental evidence for them and to study their
characteristics in detail.

Motion-induced radiation can be interpreted as a result of dephasing of
vacuum fields depending on the mirrors motion. The order of magnitude of the
dephasing is expected to be the ratio between the mirror's velocity $v$ and
the speed of light $c$. For most conceivable motion of a macroscopic object,
the velocity $v$ cannot greatly exceed the sound velocity and is thereby
much slower than that of light. This is why motion-induced radiation is very
small for a single mirror oscillating in vacuum. This conclusion holds for
perfectly reflecting mirrors as well as for partly transmitting ones.

A number of works have been devoted to photon production inside the cavity
built with a pair of perfectly reflecting mirrors moving in vacuum \cite
{Dodonov89,Dodonov95,Law94,Cole96}. However no predictions can be made for
the amount of radiation emitted outside the cavity when the resonator is
treated as a closed system. In contrast the resonant enhancement is found to
be determined by the cavity finesse when the cavity is treated as an open
system \cite{Jaekel91} from which the photons can escape. Motion induced
radiation, that is photon emission outside a cavity
oscillating in vacuum is resonantly enhanced by the cavity finesse when compared
to the radiation from a single oscillating mirror. The resonant enhancement occurs when
the mechanical frequency is a multiple of the lowest cavity mode. Even and odd multiples
correspond respectively to breathing modes, where the mechanical cavity
length changes periodically, and to translation modes, where the cavity moves as a 
whole \cite{Lambrecht96}. The latter effect reminds radiation from a single mirror 
inasmuch as vacuum fluctuations are the only reference for the cavity motion. However
the order of magnitude of photon emission may greatly exceed the one from a
single mirror. From an experimental point of view the cavity is so far the
most interesting system to look for an experimental observation of
dissipative effects of vacuum fluctuations.

Inside a cavity the field undergoes many reflections before leaving the
cavity through one of the mirrors. The number of round-trips of the field is
roughly given by the cavity finesse. In loose terms, one may define an
effective velocity where the physical velocity $v$ normalized by the speed
of light is multiplied by the number of round-trips inside the cavity.
Effective velocity and thus motion-induced radiation become the larger the
higher is the cavity finesse. The effective velocity is no longer a material
velocity and may therefore approach the speed of light. In contrast to the single
mirror's case,
qualitatively new effects are expected, such as the formation of a pulse
bouncing back and forth in the cavity \cite{Law94,Cole96}. Since the pulse
duration is shorter than the time of flight through the cavity, the
radiation spectrum should also contain various frequencies corresponding to
higher-order cavity modes and thus exceeding the mechanical 
frequency.

These effects cannot be obtained with a linear treatment as the one used
in \cite{Lambrecht96}. In such a treatment the field scattering is supposed to be
linear in the mirrors motion. 
The field-mirror interaction corresponds to a
coupling of the vacuum field radiation pressure, which is quadratic in the
field, to the mirrors mechanical motion. Photons are thus emitted in
pairs. In the linear regime the generation of motion-induced radiation is analogous
to a parametric process during which the
mechanical excitation is transformed into a pair of photons. Due to energy
conservation the sum of their frequencies equals the oscillation frequency.
Therefore motion-induced photons are only emitted at frequencies smaller than the
excitation frequency. 
The linear assumption is correct as long as the
total field dephasing due to interaction with the moving reflector remains
small. The field dephasing from one reflection scales with the mirrors
velocity over the speed of light. The linear assumption is 
always satisfied for a single macroscopic mirror.

However for a cavity the crucial parameter is the effective velocity and the total field 
dephasing can become important for a high finesse 
cavity. As a consequence we expect frequency multiplication to occur which generates 
frequencies larger than the mechanical excitation frequency. 
The linear treatment, which predicts the emission frequencies to be smaller than the 
oscillation frequency,
then looses its validity and has to be replaced by a treatment which fully
accounts for a large field dephasing produced through successive reflections
of the field onto the mirrors. This treatment will be called non-linear in
the following although the scattering is still linear in the field.

The aim of the present paper is to give a treatment of the radiation emitted
by a cavity moving in vacuum which takes into account both the effect of
accumulated field dephasing and the open character of the cavity. In
particular we will study the explicit dependence of motional radiation on
the two experimentally accessible parameters, the mirrors velocity and the
cavity finesse. We will first introduce general calculation techniques for a
single mirror and for a cavity moving in vacuum with an arbitrary motion. We
will then concentrate on a particular motion, the harmonic oscillation of
the mirrors, which allows to evaluate in closed analytical form the
correlation functions of the radiation through a special parametrization of
the motion. We will give expressions for the time-dependent radiated energy
as well as the frequency-dependent radiation spectrum.

\section{Single mirror moving in vacuum}

Neglecting all effects related to polarizations, the electromagnetic field $%
\Phi $ is considered as a scalar function of one space variable, $t$ and $x,$
and will be written as a sum of two counterpropagating components $\varphi $
and $\psi $ which are functions of two light-cone variables $u$ and $v$ 
\begin{eqnarray}
\Phi (t,x) &=&\varphi (u)+\psi (v)  \nonumber \\
u=t-x &\qquad &v=t+x
\end{eqnarray}
For the sake of simplicity, the velocity of light is set to unity. We limit
ourselves here to two-di\-men\-sio\-nal space-time calculations. As is well
known from the analysis of squeezing experiments \cite{OPO}, the transverse
structure of the cavity modes does not change appreciably the results
obtained from this simplified model. Each transverse mode is correctly
described by a two-dimensional model as soon as the size of the mirrors is
larger than the spot size associated with the mode. The two-dimensional
model thus corresponds to a situation where one transverse mode is
efficiently coupled to the moving mirrors.

We now represent the mirror's motion either by a function $x=q(t)$
associated with the trajectory or as a monotonous function $v=V(u)$ relating the light
cone variables $u$ and $v$ of the light rays intersecting on the mirror's
trajectory. A propagation component $\varphi _{{\rm out}}$ of the output
field can then be written as a function of the input fields $\varphi _{{\rm %
in}}$ and $\psi _{{\rm in}}$ and of the function $V$ 
\begin{eqnarray}
\varphi _{{\rm out}} &=&\sqrt{R}\psi _{{\rm in}}\circ V+\sqrt{T} \varphi _{%
{\rm in}}  \nonumber \\
T &=&1-R  \nonumber \\
(h\circ g)\left( u\right) &=&h\left( g\left( u\right) \right)  \label{out}
\end{eqnarray}
The symbol $\circ $ represents the composition law for functions. The
coefficients $\sqrt{R}$ and $\sqrt{T}$ are the reflection and transmission
amplitudes describing scattering upon the mirror. For the sake of
simplicity, we have assumed these coefficients to be real and frequency
independent.

We now recall the standard calculation of the energy radiated into vacuum by
the moving mirror \cite{FullingDavies76}. The input fields are supposed to
be in the vacuum state and characterized by the correlation function 
\begin{equation}
\langle \varphi _{{\rm in}}(u)\varphi _{{\rm in}}(\bar{u})\rangle = -\frac{%
\hbar }{4\pi }\ln |u-\bar{u}|-\frac{i\hbar }{8}\varepsilon (u-\bar{u})
\label{phivac}
\end{equation}
The first term corresponds to the anticommutator and is state-dependent
whereas the second term gives the commutator contribution and is
state-independent. Since $\varepsilon $ is the sign function, it is clear
that the field commutator remains unchanged under the transformation (\ref
{out}) where $u$ is replaced by a monotonous function $V\left( u\right) $.
The change of the correlation function between the input and output field is
given by the following function which depends on the field anticommutators
only 
\begin{eqnarray}
C(u,\bar{u}) &=&\langle \varphi _{{\rm out}}^{\prime }(u)\varphi _{{\rm out}
}^{\prime }(\bar{u})\rangle -\langle \varphi _{{\rm in}}^{\prime }(u)\varphi
_{{\rm in}}^{\prime }(\bar{u})\rangle  \nonumber \\
&=&-\frac{\hbar R}{4\pi }\left( \frac{V^{\prime }(u)V^{\prime }(\bar{u})} {%
(V(u)-V(\bar{u}))^{2}}-\frac{1}{(u-\bar{u})^{2}}\right)  \label{corr}
\end{eqnarray}
Throughout the paper, the prime signifies a derivative of a function with
respect to its argument. The energy density $e_{u}(u)$ radiated per unit
time is given by the function $C(u,\bar{u}=u)$ evaluated at coinciding
points through a point splitting regularization procedure \cite
{FullingDavies76} 
\begin{eqnarray}
e_{u}(u) &=&C(u,u)=-\frac{\hbar R}{24\pi }{\cal S}V\left( u\right)  \nonumber
\\
{\cal S}V &=&\frac{V^{\prime \prime \prime }}{V^{\prime }}-\frac{3}{2}\left( 
\frac{V^{\prime \prime }}{V^{\prime }}\right) ^{2}  \label{Schwartz}
\end{eqnarray}
The function ${\cal S}V$ is the Schwarzian derivative of $V$. No radiation
is emitted when the reflector has a uniform acceleration, which corresponds
to a vanishing Schwarzian derivative ${\cal S}V$. The total energy radiated
by the moving mirror can then be obtained by integrating the energy density
over $u$. In the following, we will concentrate on the particular case of an
oscillatory motion of mechanical frequency $\Omega $. In this case the
energy $E_u$ radiated per period is read as 
\begin{equation}
E_{u}=\displaystyle \int_{0}^{\frac{2\pi }{\Omega }}e_{u}(u){\rm d}u
\label{Etot}
\end{equation}

In order to characterize the radiation we have also to describe its spectral
properties. The radiation spectrum may be represented as a density of
photons obtained from the Fourier transform of the two point function $C(u,%
\bar{u})$. We will turn to its description later on.

\section{Cavity moving in vacuum}

The vacuum field is defined on both sides of the cavity as in the previous
section. The relation between input and output fields is a generalization of
(\ref{out}) which corresponds to the standard Fabry-Perot theory 
\begin{eqnarray}
\varphi _{{\rm out}} &=&-\sqrt{R_{2}}\psi _{{\rm in}}\circ V_{-1}+\sqrt{R_{1}%
}T_{2}\sum_{n\geq 0}r^{n}\psi _{{\rm in}}\circ V_{n}  \nonumber \\
&+&\sqrt{T_{1}}\sqrt{T_{2}}\sum_{n\geq 0}r^{n}\varphi _{{\rm in}}\circ U_{n}
\nonumber \\
T_{1} &=&1-R_{1}\qquad T_{2}=1-R_{2}  \nonumber \\
r &=&\sqrt{R_{1}}\sqrt{R_{2}}=e^{-2\rho }\quad  \label{phiout}
\end{eqnarray}
The reflection and transmission amplitudes of the two mirrors are related
through unitarity conditions. The coefficient $r$ determines the attenuation
factor of the field on a single cavity round-trip. It can also be written as
a function of the cavity losses $\rho $. Throughout the paper we will use $%
\rho $ when we consider the experimentally interesting case of a high
finesse cavity with $\rho \ll 1$. In the more general case the reflection
coefficient $r$ will be used. The functions $U_{n}$ and $V_{n}$ represent
the light cone variables associated with the various input rays which are
transformed into the output light ray $u$ by the cavity. When the cavity is
at rest they are given by simple relations 
\begin{eqnarray}
U_{n}\left( u\right) &=&u-2nL  \nonumber \\
V_{n}\left( u\right) &=&u-(2n+1)L  \label{unvnrest}
\end{eqnarray}
The length $L$ of the cavity is measured as a time of flight between the two
mirrors; the two mirrors are supposed to be located at $x=\pm L/2$
respectively; the ray $V_{-1}$ represents the particular case where the
field has been directly reflected back by the first encountered mirror
without entering the cavity.

\begin{figure}[h]
\centerline{\psfig{figure=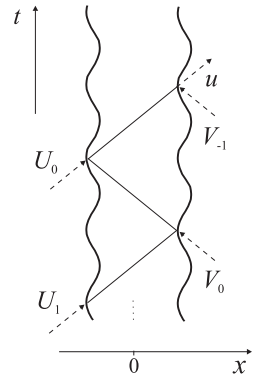,height=5cm}}
\caption{Space-time diagram of on arbitrary field trajectory bouncing back
and forth inside the cavity. Both mirrors are supposed to follow a harmonic
motion. Light rays are indicated by null lines, i.e. straight lines making a 
$45^{\circ }$-angle with the time and space axis. }
\label{figuvcav}
\end{figure}

We may deduce the field emitted by the vibrating cavity through the same
expression (\ref{phiout}) as for the motionless cavity, but with functions $%
U_{n}$ and $V_{n}$ now given by the procedure sketched on figure \ref
{figuvcav}. 
The functions $U_{n}$ and $V_{n}$ corresponding to the various light rays in
figure \ref{figuvcav} are built up through a functional iteration, 
\begin{eqnarray}
U_{n}=f_{2n} &\qquad &V_{n}=f_{2n+1}  \nonumber \\
f_{-1}=g &\qquad &f_{0}=I  \nonumber \\
f_{2n}=g^{-1}\circ f_{2n-1} &\qquad &f_{2n+1}=h\circ f_{2n}  \label{chaine}
\end{eqnarray}
where $I$ is the identity function ($I\left( u\right) =u$), while the two
functions $h$ and $g^{-1}$(inverse of $g$) represent the trajectories of the
mirrors 
\begin{eqnarray}
x &=&q_{1}(t)\Longrightarrow v=h(u)  \nonumber \\
x &=&q_{2}(t)\Longrightarrow v=g(u)\Rightarrow u=g^{-1}(v)  \label{def2mir}
\end{eqnarray}
The function $f_{2n}$ results from $n$ successive compositions of the
function $g^{-1}\circ h$. This construction is quite analogous to
the one described in \cite{Cole96}, but it is written here such that it may
be applied to an open cavity.

In a linear treatment \cite{Lambrecht96} the total
field dephasing remains small. At every reflection the field
acquires a dephasing due to the mirrors motion $2q_{i}(t),(i=1,2)$. 
In this case the composition of functions (\ref{chaine}) is reduced to
the summation of the mirrors motion. The accumulated dephasing
after $M$ roundtrips inside the cavity is now seen to be simply $M$
times the dephasing due to a single roundtrip. The scattered field then
has a temporal variation which reproduces the mirrors motion. With these
approximations the results of the linear treatment are recovered.

However, if the field undergoes a great number of roundtrips inside the
cavity, the total dephasing does not remain small and a linearization is not 
valid anymore. The dephasing has then to be calculated through the general
composition law  (\ref{chaine}). Following the same procedure as for
the single mirror, we calculate the density of energy $e_{u}(u)$ radiated to
the right through the two-point correlation function defined as in (\ref{corr}) by
letting $\bar{u}$ come to coincidence with $u$. As all functions now depend
on a single parameter $u$, we omit this parameter in the expression for the
energy density 
\begin{eqnarray}
e_{u}=- &&\frac{\hbar }{4\pi }\left\{ \frac{R_{2}}{6}{\cal S}%
f_{-1}-2T_{2}\sum_{n\geq 0}r^{n+1}\frac{f_{-1}^{\prime }f_{2n+1}^{\prime }}{%
(f_{-1}-f_{2n+1})^{2}}\right.   \nonumber \\
&+&\frac{T_{2}^{2}R_{1}}{6}\sum_{n\geq 0}r^{2n}{\cal S}f_{2n+1}+\frac{%
T_{1}T_{2}}{6}\sum_{n\geq 0}r^{2n}{\cal S}f_{2n}  \nonumber \\
&+&T_{1}T_{2}\sum_{n\neq m\geq 0}r^{n+m}\frac{f_{2n}^{\prime }f_{2m}^{\prime
}}{(f_{2n}-f_{2m})^{2}}  \nonumber \\
&+&\left. T_{2}^{2}R_{1}\sum_{n\neq m\geq 0}r^{n+m}\frac{f_{2n+1}^{\prime
}f_{2m+1}^{\prime }}{(f_{2n+1}-f_{2m+1})^{2}}\right\}   \label{ecav1}
\end{eqnarray}
Compared to the radiated energy density of a single mirror (\ref{Schwartz}),
we now find a sum of Schwarzian derivatives corresponding to different
numbers $n$ of round-trips inside the cavity, as well as new terms arising
from the interference between light rays having undergone a different number
of roundtrips.

The derivatives appearing in the upper equation are iteratively deduced from
each other through relations (\ref{chaine}) and the chain rules associated
with derivation of composed functions 
\begin{eqnarray}
\left( g\circ h\right) ^{\prime } &=&h^{\prime }\ g^{\prime }\circ h 
\nonumber \\
{\cal S}\left( g\circ h\right) &=&{\cal S}h+h^{\prime }{}^{2}\left( {\cal S}
g\right) \circ h  \label{chainfg}
\end{eqnarray}
In the general case of arbitrary motions of the two mirrors, the various
relations which have been written in the present section allow to
compute the energy density radiated by a cavity built with partly
transmitting mirrors.

\section{Harmonic motions and periodic orbits}

{}From now on we focus our attention on configurations which have been shown
to be the most efficient ones to generate motion-induced radiation \cite
{Lambrecht96} and which furthermore allow us to put the problem in a simpler
form.

We consider that the two mirrors follow harmonic motions at such a frequency
that the motion-induced effects, i.e. motional radiation and motional force,
are resonantly enhanced by the multiple interference taking place inside the
cavity. The frequency $\Omega $ of the harmonic motion is thus supposed to
be such that $\Omega L$ is a multiple of $\pi $ 
\begin{equation}
\Omega L=K\pi
\end{equation}
The amplitudes of the two motions are supposed to have the same absolute
value with either opposite or identical signs depending on the parity of the
integer number $K$ 
\begin{eqnarray}
\Omega q_{1}(t) &=&-\frac{K\pi }{2}-\beta \sin \left( \Omega t- \frac{%
(K+1)\pi }{2}\right)  \nonumber \\
\Omega q_{2}(t) &=&\frac{K\pi }{2}-\beta \sin \left( \Omega t+\frac{(K+1)\pi 
}{2}\right)  \nonumber \\
\beta &=&{\rm th}(\alpha )  \label{sinmotion}
\end{eqnarray}
We have written the two equations of motion in terms of dimensionless
numbers. In particular, one distinguishes two parameters $\alpha $ and $%
\beta $. $\beta $ represents the ratio between the maximal velocity of the
mirrors and the velocity of light while $\alpha $ plays the role of a
rapidity which will be found to add up through successive reflections when
one considers the composed motion of both mirrors.

When $K$ is an even number, the upper equations describe a situation where
the two mirrors are oscillating such that the length of the cavity changes
periodically. In the opposite case, the cavity performs a global oscillation
with its length kept constant. The two cases will be called even and odd
modes respectively in the following. A cavity motion corresponding to an odd
mode is reminiscent of motion-induced radiation from a single oscillating
mirror, but in addition here the radiation is resonantly enhanced inside the
cavity. These statements follow from the linearized approach \cite
{Lambrecht96} but we expect a similar behavior to take place in the full
treatment developed in the present paper.

There exist periodic orbits such that the optical length seen by the field
bouncing back and forth in the cavity is the same on successive round-trips
despite the motion of the mirrors \cite{Cole96}. These orbits correspond to
particular values $\tilde{u}$ of the light-cone variable $u$ such that the
iteration procedure leads to expressions similar to the ones obtained when
the mirrors are at rest (cf. (\ref{unvnrest})). Although definition (\ref
{chaine}) of $f_{p}$ is different depending on whether $p$ is even or odd,
the periodic orbits generalize the usual resonance condition of the
Fabry-P\'{e}rot theory 
\begin{equation}
f_{p}(\tilde{u})=\tilde{u}-pL
\end{equation}
They therefore give rise to a constructive interference effect, analogous to
that occurring for a motionless cavity. Since the light rays corresponding to
a periodic orbit encounter the mirrors at the same position after an
arbitrary number of round-trips, the composition (\ref{chaine}) of motions
leads to a simple power law for the derivatives evaluated after $n$
roundtrips as well as for the Schwarzian derivatives 
\begin{eqnarray}
f_{p}^{\prime }(\tilde{u}) &=&e^{2p\alpha }  \nonumber \\
{\cal S}f_{p}(\tilde{u}) &=&{\cal S}f(\tilde{u})\frac{1-e^{4p\alpha }} {%
1-e^{4\alpha }}  \label{Scav}
\end{eqnarray}
There exist two sets of periodic orbits which correspond to opposite values
of $\alpha $ in (\ref{Scav}). These two periodic orbits attract (respectively repel) 
the neighboring trajectories, when $e^{4\alpha }$ is greater (respectively smaller) 
than unity. Only the
attractive orbit is expected to give rise to a large enhancement of the
motional radiation.

In expression (\ref{ecav1}) of the energy density the Schwarzian derivative (%
\ref{Scav}) is multiplied by the squared reflection coefficient after $n$
roundtrips $r^{2n}$. Summation over the number of roundtrips leads then to a
geometric progression of $r^{2n}e^{8n\alpha }$. The first factor represents
the attenuation of the energy density associated with the cavity losses
through the two mirrors. The second one accounts for the parametric
amplification of the field associated with the mirrors' motion. As a
consequence, the energy density takes large values when the parametric
amplification compensates the losses. In fact, a divergence of the energy
density should occur when $re^{4\alpha }$ approaches unity. This corresponds
exactly to the oscillation threshold of a mechanically excited parametric
amplifier. Let us notice that the approach developed in the present paper
does not remain valid above this threshold.

We have focussed our attention here on the case where periodic orbits
correspond to light rays meeting the two mirrors at their mean positions,
respectively $-\frac L 2$ and $\frac L 2$. There exist more general
situations where the light rays meet the mirrors at other positions \cite
{Cole96} which will not be considered here. Notice that the
particular case studied in the present paper is interesting from an
experimental point of view since it corresponds to the maximum value of the
parameter $\alpha$ for motions having a given frequency and a given
amplitude.

In the following, we will restrict our attention to the cases of practical
interest where the physical velocity of the mirror remains small when
compared to the velocity of light. As discussed in the Introduction, this
condition is always met for macroscopic mirrors. It implies that
a single reflection produces a small dephasing on the field and, thereby,
small radiation effects. Precisely, this means that the quantity ${{\cal S}f}
$ which appears in (\ref{Scav}) has an extremely small value while the
factor $e^{2\alpha }$ is very close to unity. It follows that a large number 
$n$ of roundtrips is needed to obtain a factor $e^{2n\alpha }$ differing
appreciably from unity and therefore giving rise to a noticeable radiation.
We show now that this assumption permits to perform the functional iteration
in an analytical manner.

The crucial point is that the functional iteration (\ref{chaine}) may in
this case be restricted to the sub-space of periodic functions $h$
corresponding to homographic relations between the phases $e^{i\Omega u}$
and $e^{i\Omega h(u)}$ 
\begin{equation}
e^{i\Omega h(u)}=\frac{ae^{i\Omega u}+b}{b^{*}e^{i\Omega u}+a^{*}}
\Longrightarrow A(h)=\left( 
\begin{array}{ll}
a & b \\ 
b^{*} & a^{*}
\end{array}
\right)  \label{homo1}
\end{equation}
$a$ and $b$ are two complex constants and $a^{*}$ and $b^{*}$ their complex
conjugates which can be gathered in a matrix $A(h)$ associated with the
function $h$. Attention may be restricted to matrices having a determinant
equal to unity. In the sub-space of functions (\ref{homo1}), the composition
of two functions merely corresponds to a product of their corresponding
matrices. 
\begin{equation}
A(h\circ g)=A(h)A(g)
\end{equation}

Rigorously speaking, the function (\ref{homo1}) does not correspond to
sinusoidal trajectories (\ref{sinmotion}) but rather to specific
trajectories already considered by Law for dealing with photon production
inside a closed cavity \cite{Law94} 
\begin{eqnarray}
a=e^{i\phi _{a}}\text{ch}\alpha &\qquad &b=e^{i\phi _{b}}\text{sh}\alpha 
\nonumber \\
\sin \left( \Omega q-\phi _{a}\right) &=&-\beta \sin \left( \Omega t-\phi
_{b}\right)  \label{hmotion}
\end{eqnarray}
where the reduced velocity $\beta $ gives the mirrors velocity compared to
the speed of light. Variations of the phase factors $\phi _{a}$ and $\phi
_{b}$ amount to displacements of the trajectory along the space and time
axis. At the limit of small velocities however, the trajectory (\ref{hmotion}%
) is reduced to an ordinary sinusoidal motion (\ref{sinmotion}) 
\begin{equation}
\beta \ll 1\Longrightarrow \Omega q=\phi _{a}-\beta \sin \left( \Omega
t-\phi _{b}\right)
\end{equation}
The difference between the two motions scales as the cube $\beta ^{3}$ of
velocity and is therefore extremely small for realistic motions of
macroscopic mirrors. The effect of the trajectory (\ref{hmotion}) is thus
indistinguishable from the effect of the sinusoidal motion for a single
reflection.

If one considers two oscillating mirrors, the two functions $h$ and $g,$
corresponding to the first and second mirror, are associated with two
matrices $A(h)$ and $A(g)$ respectively. The matrix components are chosen to
fit equations (\ref{sinmotion}) of motion of the two mirrors 
\begin{eqnarray}
A(h) &=&\left( 
\begin{array}{ll}
(-i)^{K}{\rm ch}\alpha & i^{K+1}{\rm sh}\alpha \\ 
(-i)^{K+1}{\rm sh}\alpha & i^{K}{\rm ch}\alpha
\end{array}
\right) \quad  \nonumber \\
A(g) &=&\left( 
\begin{array}{ll}
i^{K}{\rm ch}\alpha & (-i)^{K+1}{\rm sh}\alpha \\ 
i^{K+1}{\rm sh}\alpha & (-i)^{K}{\rm ch}\alpha
\end{array}
\right)
\end{eqnarray}
As the composition of the two functions corresponds to a product of their
matrices, the composition law naturally produces a homographic function when
the number of reflections becomes large. This essential feature will be used
in the following to compute the temporal and spectral characteristics of
motional radiation from purely algebraic manipulations of the associated
matrices. In particular, the functions $f_{p}$ in equation (\ref{chaine})
corresponding to successive reflections of the field inside the cavity are
obtained through matrix multiplication 
\begin{eqnarray}
A_{p} &=&\left( 
\begin{array}{ll}
(-i)^{Kp}{\rm ch}p\alpha & i^{2K+1}(-i)^{Kp}{\rm sh}p\alpha \label{homo} \\ 
(-i)^{2K+1}i^{Kp}{\rm sh}p\alpha & i^{Kp}{\rm ch}p\alpha
\end{array}
\right)
\end{eqnarray}
The discrepancy between the composed functions built on the two motions (\ref
{sinmotion}) and (\ref{hmotion}) does not affect the results if $\beta \ll 1$%
. More precisely, the difference between the composed functions $f_{p}$
built on the two motions (\ref{sinmotion}) and (\ref{hmotion}) remains of
the order of $\beta ^{2}$ when the number of iterations increases.

\section{Pulse shaping}

In the following, we will analyze the case of a single mirror following a
trajectory (\ref{homo1}) with an arbitrary velocity parameter $\alpha $.
Although the hypothesis of a large rapidity $\alpha$ is not realistic for a
single mirror, it can be used as a model for the composition of a large
number of round trips inside the cavity. We will then come to the full
treatment of the cavity where interferences have also to be accounted for.

The derivative $h^{\prime }(u)$ may be written from (\ref{homo1}) as 
\begin{equation}
h^{\prime }(u)=\frac{1}{\left| a+be^{-i\Omega u}\right| ^{2}}
\end{equation}
$h^{\prime }(u)$ oscillates between the extremal values $\exp \left( \pm
2\alpha \right) $ which correspond to physical velocities of the mirror $\pm
\beta $. For homographic functions (\ref{homo1}), the Schwarzian derivative
has the simple form 
\begin{equation}
{\cal S}h=\frac{\Omega ^{2}}{2}\left( 1-h^{\prime 2}\right)
\label{Schwartz2}
\end{equation}
The total energy radiated to the right by a single mirror is then obtained
by averaging the energy density (\ref{Schwartz}) over one oscillation period 
\begin{equation}
E_{u}=\frac{\hbar R\Omega }{12}{\rm sh}^{2}\alpha  \label{Efin}
\end{equation}
This energy does not saturate when the parameter $\alpha $ increases
although the velocity scales as th$\alpha $ and remains smaller than the
velocity of light. The radiated energy is always greater than the squared
velocity th$^{2}\alpha $ which was the value suggested by the linear
treatment \cite{Lambrecht96}. However it is impossible to obtain appreciable
radiation with a single oscillating mirror for velocities small compared to
the speed of light. In the realistic case of a mirror moving at a small
velocity the radiated energy as well as the number of emitted photons scale
with $\alpha ^{2}$.

We come now to the energy density radiated by an oscillating cavity,
assuming that a large number of round-trips is necessary to compensate the
small velocity of the mirrors and thus get a noticeable radiation. The
energy density $e_{u}$ may be obtained from (\ref{ecav1}) by using the
following properties of the Schwarzian derivative of $f_p$ and of the
first-order derivative $f_p^{\prime}$ respectively 
\begin{eqnarray}
{\cal S}f_p&=&\frac{\Omega ^{2}}{2}\left( 1-f_p^{\prime 2}\right)  \nonumber
\\
f_{p}^{\prime }&=&\frac{1}{{\rm ch}2p\alpha +(-1)^{K}{\rm sh}2p\alpha \sin
\Omega u}  \label{Schwartz3}
\end{eqnarray}
To plot the energy density for different linear and non-linear regimes we
introduce effective quantities 
\begin{eqnarray}
\alpha _{{\rm eff}}&=&2\alpha /\rho  \nonumber \\
\beta _{{\rm eff}}&=&{\rm th}(\alpha _{{\rm eff}})
\end{eqnarray}
The effective rapidity $\alpha _{{\rm eff}}$ is given by the roundtrip value
of the rapidity multiplied by the cavity finesse $\rho ^{-1}$. In contrast
to the mechanical velocity $v$ normalized by the speed of light, which has
to remain much smaller than 1 in any physical situation, the corresponding
effective velocity $\beta _{{\rm eff}}$ can become an important fraction of
the speed of light when the field undergoes a large number of reflections
inside the cavity. The maximal value of the effective reduced velocity is
limited to $\beta _{{\rm eff}}\sim 0.76$ by the divergence of the energy
density at $\alpha _{{\rm eff}}=1$. This value corresponds indeed to the
threshold $r e^{4\alpha}=1$ which has already been mentioned previously for
the periodic orbits.

\begin{figure}[htb]
\centerline{\psfig{figure=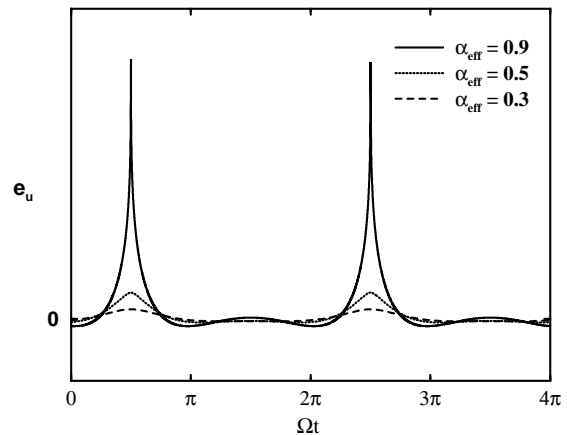,height=6.5cm}}
\caption{Energy density emitted to the right by the cavity as a function of
time for different effective rapidities $\alpha _{{\rm eff}}$ and $r=0.99$. 
The top line
of the frame corresponds to $10^{-3}\hbar \Omega ^{2}$. With increasing
values of the effective rapidity the energy starts to concentrate in pulses
emitted periodically by the cavity. The pulses become very sharp close to
the threshold of oscillation $\alpha _{{\rm eff}}=1.$}
\label{figEu}
\end{figure}

The variation of the energy density for different parameters $\alpha _{{\rm %
eff}}$ is presented on figure \ref{figEu}. In the linear regime where $%
\alpha _{{\rm eff}}\ll 1$ the temporal variation of the emitted energy is
sinusoidal. When $\alpha _{{\rm eff}}$ increases the energy concentrates in
pulses which are periodically emitted by the cavity. This pulse shaping
becomes the more pronounced, the width of the pulses the smaller, the larger
becomes the effective rapidity and thus the accumulated field dephasing.

In the same manner as for the single mirror the total energy radiated by the
cavity is computed by averaging the energy density over one oscillation
period $2\pi /\Omega $. 

\noindent As previously we will restrict our attention to a realistic case,
where the mirrors velocity is small which justifies the use of the
homographic relations. In this case we may replace the differences of
functions in the denominators of (\ref{ecav1}) by their motionless values (%
\ref{unvnrest}). Higher order contributions decrease with the inverse of
their squared order. The sums over the number of roundtrips can then be
performed and we find the following expression for the energy radiated to
the right 
\begin{eqnarray}
E_{u} &=&\frac{\hbar \Omega R_{2}}{12}{\rm sh}^{2}\alpha +\frac{\hbar \Omega
T_{2}}{48}\left( \zeta _{u}(\alpha )+\zeta _{u}(-\alpha )-2\right)  
\nonumber \\
&-&\frac{\hbar \Omega T_{2}}{8\pi ^{2}K^{2}}\left( \xi (\alpha )(\zeta
_{u}(\alpha )-e^{-2\alpha })+\xi (-\alpha )(\zeta _{u}(-\alpha )-e^{2\alpha
})\right)   \nonumber \\
\zeta _{u}(\alpha ) &=&\frac{(1-e^{-4\rho })e^{2\alpha }+T_{1}(1-e^{2\alpha
})}{1-e^{4(\alpha -\rho )}}  \nonumber \\
\xi (\alpha ) &=&\sum_{l=1}^{\infty }\frac{e^{2l(\alpha -\rho )}}{l^{2}{\rm %
ch}2\alpha l}  \label{Etotcav}
\end{eqnarray}
The energy $E_{v}$ radiated to the left can be obtained from the above
formula by interchanging the indices $1$ and $2$ of the reflection and
transmission coefficients. The total energy radiated in both directions is
then evaluated as the sum of the two contributions 
\begin{equation}
E=E_{u}+E_{v}  \label{Euv}
\end{equation}
The total radiated energy would diverge for $\alpha =\rho $. However this
limit is not reached as the energy density already diverges when $2\alpha
/\rho $ approaches unity. The reason for this difference is simply that the
widths of the various contributions to the energy density decrease when the
number of roundtrips $n$ increases, so that the contribution to the
integrated energy increases less rapidly than the peak value of the energy
density. As a consequence, the divergence of the peak density occurs before
the divergence of the integrated energy. We remind here that our calculation
does not remain valid above the oscillation threshold. Still it may be
expected that a large amount of radiation is emitted above threshold.

For experimental reasons one might also be interested in the amount of
energy stored inside the cavity. Let us first remark that inside the cavity
the fields propagating to the right and to the left are not independent from
each other due to the boundary conditions. All intracavity quantities thus
contain the field components of both directions and are obtained as the sum
of the two contributions. Having this argument in mind the intracavity
energy, integrated over the cavity length $L=K\pi /\Omega $, is then with
the same notations as above found to be 
\begin{eqnarray}
{\cal E} &=&\frac{\hbar \Omega K}{48}\left( \zeta (\alpha )+\zeta (-\alpha
)-2\right)   \nonumber \\
&-&\frac{\hbar \Omega }{8\pi ^{2}K}\left( \xi (\alpha )\zeta (\alpha )+\xi
(-\alpha )\zeta (-\alpha )-2\xi (0)\right)   \nonumber \\
\zeta (\alpha ) &=&\frac{1}{2}(\zeta _{u}(\alpha )+\zeta _{v}(\alpha ))
\label{Ecav}
\end{eqnarray}
The energy density is here directly expressed with respect to the static
Casimir energy which is recovered when the cavity is motionless \cite
{FullingDavies76}. This result is due to the fact that the vacuum outside
and inside the cavity is not the same but differs exactly by this amount of
energy.

In order to obtain an appreciable value for the radiated energy if the
cavity is moving at a small velocity, it is necessary to consider a high
finesse cavity $\rho \ll 1$, keeping in mind that the finesse should be
limited by the condition $\alpha_{{\rm eff}} < 1$. Using these assumptions
equations (\ref{Euv},\ref{Ecav}) may be approximated as follows by expanding
separately the common denominator and numerators 
\begin{eqnarray}
E &\approx &\frac{\hbar \Omega }{6}\alpha ^{2}+\frac{\hbar \Omega }{6}\left(
1-\frac{1}{K^{2}}\right)\frac{\rho \alpha ^{2}}{\rho ^{2}-\alpha ^{2}} 
\nonumber \\
{\cal E} &\approx& \frac{\hbar \Omega}{24} \left (K - \frac{1}{K} \right ) 
\frac{\alpha^2}{\rho ^{2}-\alpha ^{2}}  \nonumber \\
\alpha &\le& \frac{\rho}{2} \ll 1  \label{Eapprox}
\end{eqnarray}
The first term in the radiated energy is due to the field which is directly
reflected by the two mirrors without entering the cavity. This term
corresponds to the expression for motion-induced radiation from a single
perfectly reflecting mirror. The second term has its origin in the field
which has traversed the cavity and thus accumulated a much more important
dephasing than the singly reflected field. Neglecting $\alpha ^{2}$ in the
denominator leads to the linear result presented in \cite{Lambrecht96}. The
linear approximation is found to be rigorously valid if the rapidity is much
smaller than the cavity losses ($\alpha \ll \rho$). Furthermore the present
treatment allows us to calculate motion-induced radiation when the field
dephasing becomes large due to accumulation on a large number of reflections.
Equations (\ref{Eapprox}) have a range of validity extending up to the
threshold $\alpha_{{\rm eff}}=1$.

The intracavity energy contains only the term corresponding to the field
which has entered the cavity. Its expression can also be deduced from the
radiated energy through a detailed balance argument, which goes as follows 
\cite{Lambrecht96}: The energy inside the cavity can be obtained from the
radiated energy by multiplying it by two factors $\Omega /(2\pi )$ and $%
2L/(4\rho )$. The first factor is due to the fact that the radiated energy
is the energy density integrated over one oscillation period. During one
roundtrip of duration $2L$ each photon has the probability $4\rho $ -
corresponding to the energy transmission coefficient of the two mirrors - to
escape from the cavity. The present non-linear evaluation of the intracavity
energy gives indeed the same result as the balance argument in the limiting
case of a high finesse cavity (cf. equations (\ref{Eapprox})). However, as equations 
(\ref{Etotcav},\ref{Ecav}) show the balance argument is not true for a cavity
with arbitrary reflection and transmission coefficients. A remarkable
consequence of equations (\ref{Eapprox}) is that the non-linear calculation
is necessary as soon as the number of photons inside the cavity becomes of
the order of unity.

Interesting remarks can be made concerning the particular case $K=1$. 
Clearly equations (\ref{Eapprox}) show that for a high finesse cavity no
enhancement of photon production inside the cavity can be obtained when the
mechanical excitation frequency equals the lowest cavity mode ($K=1$)
in accordance with results in ref. \cite{Dodonov95,Law94}. In this case the
energy $E$ radiated by the cavity corresponds to the one emitted by a
single mirror and the motional intracavity energy vanishes. However, in the
general case of arbitrary cavity finesse motion-induced photons are also
found for the $K=1$ mode (cf. equation (\ref{Ecav})). 
The key point is that as far as classical light rays 
are concerned the mode $K=1$ behaves like all other modes \cite{Cole96}. 
However the field dephasing and thus motion-induced
radiation is not only determined by the behavior of the light rays but also
by the cavity which plays the part of a filtering function and suppresses photons 
at zero frequency. As a consequence of the coupling to radiation pressure
photons are not emitted singly but in pairs. Thus motion-induced
radiation is enhanced by the cavity if all photons are emitted into a
cavity mode, the sum of their frequencies being equal to the
mechanical oscillation frequency. In order to fulfill this condition when
the cavity oscillates with the frequency of the lowest cavity mode 
photons have to be emitted at zero frequency. The cavity suppresses those
photons the more efficiently the higher is the cavity finesse. Thus
motion-induced photons for the $K=1$ mode can be found in the bad
cavity limit but not in the high finesse limit.

Coming back to the general case, we emphasize that equations (\ref{Eapprox}) remain 
valid up to the threshold $\rho =2\alpha $,
when the cavity finesse $\rho ^{-1}$ is increased with the amplitude of
motion kept constant. Below this value motion-induced radiation is amplified inside the
cavity, but the cavity losses exceed the amplification gain. As expressions (%
\ref{Eapprox}) are monotonous in $\rho $ their maximum values are thus
reached at threshold. In this regime we then find a maximum of the radiated
energy which depends linearly on $\rho $. Comparing this value to the
maximum energy emitted by a single oscillating mirror shows a gain of the
order of the cavity finesse by considering a cavity instead of a single
mirror. The cavity is thus a much more favorable system to produce
motion-induced radiation than a single mirror. Furthermore if one increases
the cavity finesse above its threshold value the roundtrip amplification of
the field due to the mirrors motion should exceed the cavity losses and the
system should enter a regime of exponential amplification. Without further
calculations we then expect the oscillating cavity to emit photon pulses of
much higher intensity above threshold than below.

\section{Frequency up-conversion}

We now turn to the calculation of the radiation spectrum where we will
proceed as previously by first studying the case of a single moving mirror
and afterwards the one of the oscillating cavity.

The scattering field equation (\ref{out}) writes in Fourier space 
\begin{eqnarray}
\varphi _{{\rm out}}[\nu ] &=&\sqrt{T}\varphi _{{\rm in}}[\nu ]+\sqrt{R}\int 
{\rm d}\bar{\nu}\frac{\Omega }{2\pi }\psi_{{\rm in}}[\bar{\nu}]\int {\rm d}u
e^{i\Omega(\nu u+\bar{\nu}V(u))}  \nonumber \\
\nu &=&\frac{\omega }{\Omega }\qquad \bar{\nu}=\frac{\bar{\omega}}{\Omega }
\end{eqnarray}
where we have introduced the reduced frequencies $\bar{\nu}$ and $\nu $
normalized with respect to the mechanical frequency $\Omega .$ The field
dephasing of the output field is determined by $V(u)$ and thus associated
with the mirrors position $Q(u)$ which is easily calculated from (\ref{homo1}%
) 
\begin{equation}
\Omega Q\left( u\right) =\Omega \frac{V(u)-u}{2}={\rm arctg}\left( \frac{%
\beta \cos \Omega u}{1+\beta \sin \Omega u}\right)
\end{equation}
$e^{2i\bar{\nu}Q(u)}$ is a periodic function and can thus be developed into
a Fourier series with discrete coefficients $\gamma _{m}$ 
\begin{equation}
e^{2i\bar{\nu}\Omega Q(u)}=\sum_{m}\gamma _{m}[\bar{\nu}]e^{-im\Omega u}
\label{Fourier}
\end{equation}
which will determine the radiation spectrum. If the mirrors motion were
sinusoidal these Fourier coefficients would be given by Bessel functions of
different orders. We have now to evaluate these coefficients for a
homographic trajectory (\ref{hmotion}).

To this aim, we first write the field dephasing (\ref{Fourier}) as 
\begin{equation}
e^{2i\Omega \bar{\nu}Q\left( u\right) }=\left( \frac{1+i\beta e^{-i\Omega u}%
} {1-i\beta e^{i\Omega u}}\right) ^{\bar{\nu}}  \label{fielddephas}
\end{equation}
The Fourier coefficients (\ref{Fourier}) may be rewritten in terms of an
hypergeometric series \footnote{%
The coefficient $G_m(\nu,\beta)$ is directly related to the hypergeometric
function $F(\nu,m-\nu;m+1;\beta^2)$ defined for instance in \cite{Grad9100}.
We have also used property 8.334 of the $\Gamma$-function in the same
reference.} 
\begin{eqnarray}
\gamma _{m}[\bar{\nu}] &=&\left( -i\right) ^{m+2}\frac{\bar{\nu}}{\pi }\sin
\left( \pi \bar{\nu}\right) G_{m}\left( \bar{\nu},\beta \right)  \nonumber \\
G_{m}\left( \nu ,\beta \right) &=&\beta ^{m}\sum_{l\geq 0}\frac{\Gamma
\left( \nu +l\right) \Gamma \left( m-\nu +l\right) }{\Gamma (m+1+l)} \frac{%
\beta ^{2l}}{l!}  \label{defG}
\end{eqnarray}

The radiation spectrum, that is the spectral density of the photon number
per unit time and defined for positive frequencies, is then given by \cite
{Lambrecht96} 
\begin{eqnarray}
n_{\nu } &=&R\sum_{m\geq \nu }\frac{\nu }{m-\nu }|\gamma _{m}[m-\nu ]|^{2} 
\nonumber \\
&=&R\frac{\sin ^{2}(\pi \nu )}{\pi ^{2}}\sum_{m>\nu }\nu \left( m-\nu
\right) \left| G_{m}\left( \nu ,\beta \right) \right| ^{2}  \label{specfinal}
\end{eqnarray}
The total energy may be recovered as the integral of the spectrum as well as
the integral of the energy density. The radiation spectrum vanishes for all
values $\nu $ equal to a natural number, that is for all frequencies $\omega 
$ equal to a multiple of $\Omega $. The spectrum thus decomposes into a
succession of arches, each limited by two successive multiples of the
excitation frequency.

The upper expression corresponds to reflection upon a single moving mirror.
In order to get the spectral distribution of radiation from the cavity, we
now have to take into account the interferences between light rays having
undergone different number of reflections inside the cavity. We proceed in
the same manner as in the case of a single moving mirror by splitting the
function $f_{p}$ into two parts. The first part is linear in the parameter $%
u $ while the second one $Q_{p}$ is induced by the motion and harmonic like
the mirrors' motion 
\begin{eqnarray}
f_{p}\left( u\right) &=&u-pL+2Q_{p}\left( u\right)  \nonumber \\
\Omega Q_{p}(u) &=&{\rm arctg}\left( \frac{\beta _{p}\cos \Omega u}{1+\beta
_{p}\sin \Omega u}\right)  \nonumber \\
\beta _{p} &=&(-1)^{K}{\rm th}\left( p\alpha \right)
\end{eqnarray}
The round-trip dephasing $2L$ corresponds to the case of periodic orbits so
that the function $Q_{p}$ vanishes at these points. The periodic function $%
e^{2i\bar{\nu}Q_{p}}$ can be developed into a Fourier series with
coefficients now depending on the number of round trips $p$. We proceed as
previously to find the Fourier coefficients 
\begin{eqnarray}
\gamma _{m,p}[\bar{\nu}] &=&\int_{0}^{\frac{2\pi }{\Omega }}\frac{\Omega } {%
2\pi }du~e^{im\Omega u}e^{iK\pi \bar{\nu}p}e^{2i\Omega \bar{\nu}Q_{p}\left(
u\right) }  \nonumber \\
&=&\left( -i\right) ^{m+2}\bar{\nu}\frac{\sin \left( \pi \bar{\nu}\right) }{%
\pi }e^{iK\pi \bar{\nu}p}G_{m}\left( \nu ,\beta _{p}\right)
\end{eqnarray}
and the radiation spectrum 
\begin{eqnarray}
n_{\nu } &=&\frac{\sin ^{2}(\pi \nu )}{\pi ^{2}}\sum_{m>\nu }\nu \left(
m-\nu \right)  \nonumber \\
&\times &\left\{ \left| \sqrt{R_{1}}T_{2}\sum_{n\geq 0}r^{n}e^{-2i\pi K\nu
(n+1)}G_{m}\left( \nu ,\beta _{2n+1}\right) \right. \right.  \nonumber \\
&-&\left. \sqrt{R_{2}}G_{m}\left( \nu ,\beta _{-1}\right) \right| ^{2} 
\nonumber \\
&&\left. +T_{1}T_{2}\left| \sum_{n\geq 0}r^{n}e^{-2i\pi K\nu n}G_{m}\left(
\nu ,\beta _{2n}\right) \right| ^{2}\right\}  \label{specfinalcav}
\end{eqnarray}
Let us mention that we recover the predictions of the linearized treatment
for a motion with a small velocity by keeping only the lowest-order term $%
m=1 $ in the hypergeometric series. The spectrum is then parabolic and found
to be restricted to the frequency range corresponding to the first arch \cite
{Lambrecht96}.

\begin{figure}[htb]
\centerline{\psfig{figure=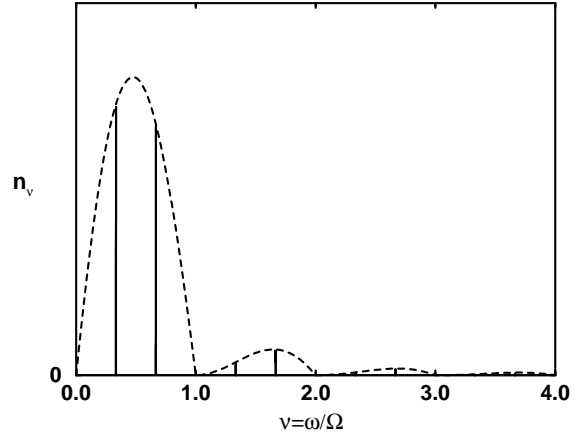,height=6.5cm}}
\caption{Spectrum of the radiation emitted by the cavity for $\alpha _{{\rm %
eff}}=0.9$ and a reflection coefficient of $r=0.99$. The top line of the
frame corresponds to $\alpha_{{\rm eff}}^2/4$. The peaks correspond to
cavity resonance frequencies. The spectrum is plotted for a cavity
oscillating globally at a mechanical frequency $\Omega =3\pi /L$. The dashed
line constitutes the envelope of the spectrum. It was obtained by averaging
the spectrum of a single mirror over the effective velocities corresponding
to different number of roundtrips. Photons are created at frequencies higher
than the mechanical oscillation frequency through frequency up-conversion in
the opto-mechanical coupling between vacuum fluctuations and the mirrors
motion. Furthermore the radiation spectrum vanishes for frequencies equal to
a multiple integer of the mechanical excitation frequency.}
\label{figspec}
\end{figure}

Figure \ref{figspec} shows the radiation spectrum for an effective rapidity $%
\alpha _{{\rm eff}}=0.9$ near the threshold of parametric oscillation. The
spectrum shown here is plotted for a cavity oscillating globally at a
frequency of $\Omega =3\pi /L$. This
means that the cavity performs three oscillations during one roundtrip of
the field inside the cavity. The dashed line was obtained by putting
formally $K=0$ in equation (\ref{specfinalcav}) which eliminates the phase
factors responsible for the interferences. It may be interpreted as the
spectrum of radiation emitted by a single oscillating mirror averaged over
the effective velocity. Clearly photons can be created by higher-order
harmonics of the motion as well as by the fundamental one as soon as the
effective velocity becomes appreciable
compared to the speed of light. As a striking consequence, photons are
radiated at frequencies higher than the mechanical frequency $\Omega $. A
process of frequency up-conversion thus exists in the opto-mechanical
coupling between vacuum fluctuations and mechanical motion of scatterers. 
\noindent A corresponding situation is found for a single oscillating mirror
(dashed line) which is however not realistic as it would imply a mirror's
mechanical velocity appreciable compared to the speed of light. The use of a
cavity allows to reproduce the same spectral density within the bandwidth of
the cavity resonance lines for realistic mirrors' velocities. A second
striking feature is that no photons are emitted at frequencies equal to
multiple integers of the excitation frequency $\Omega$ neither by a single
oscillating mirror nor by a vibrating cavity.

When comparing the cavity radiation spectrum to expression (\ref{specfinal})
corresponding to a single mirror, a difference is the emergence of peaks
typical of cavity resonances. In fact, the interferences between the pathes
corresponding to different numbers $n$ of round-trips are essentially
determined by the factors $r^{n}e^{-2in\pi K\nu }$ and $r^{n}e^{-2i\pi K\nu
(n+1)}$. The propagation dephasing after one round-trip is $e^{2i\pi K\nu }$
where $K$ is the order of the mechanical frequency as compared with the
fundamental resonance frequency of the cavity. It follows that the peaks are
apparent at frequencies equal to an integer multiple of $K^{-1}$, as shown
on Fig. \ref{figspec} with $K=3$. Their shape is Lorentzian for a high
finesse cavity. The width of each peak is given by the inverse of the cavity
finesse. The number of peaks fitting into the interval $[0,\Omega ]$
corresponds to the order $K$ of the excited cavity mode compared to the
mechanical frequency.

\section{Discussion}

In this paper we have presented a non-linear calculation of motion-induced
radiation from a cavity taking fully into account the accumulation of
dephasing through successive reflections of the field on partly transmitting
mirrors. This approach confirms the main results of the linearized treatment
which was previously used and makes it possible to specify its range of
validity. Furthermore the non-linearity due to the accumulative field
dephasing produces particular signatures of motion-induced radiation which
cannot be calculated within the linear approximation.

In the experimentally relevant case where the mirrors move with a velocity
small compared to the speed of light the emitted photon number from a single
mirror moving in vacuum scales with its squared velocity. Compared to this
situation motion-induced radiation from an oscillating cavity is enhanced by
the cavity finesse. For high finesse cavities as they exist for instance in
the microwave-regime this enhancement brings motion-induced radiation within
reach of an experimental observation. This clearly proves the cavity to be a
much more favorable system for the generation of motion-induced radiation.

In addition the present calculation shows that the linear approach is valid
when the effective rapidity, given by the mirrors' physical velocity
multiplied by the cavity finesse, is much smaller than 1. We have given here
expressions having a much larger range of validity.

In order to measure motion-induced radiation it is necessary to dispose of
signatures which permit to distinguish vacuum radiation from spurious
effects. The present calculation has allowed to identify two quantities
showing signatures which could serve to this aim, the temporal variation of
the radiated energy density and the spectral density of the emitted photon
number.

We have studied the emitted energy density as a function of different
effective rapidities. With increasing effective rapidity the energy starts
to concentrate in pulses which are emitted periodically into vacuum by the
cavity. These pulses become the higher and the sharper the more the
effective rapidity approaches its threshold value. The energy density
diverges when the single-reflection rapidity equals half of the cavity losses
during one roundtrip ($\alpha =\rho /2$). The characteristic temporal
variation which allows high energy densities in regularly spaced and narrow
time windows might be exploited in an experimental observation.

The spectrum of motion-induced radiation shows several remarkable features.
First photons may be radiated at frequencies higher than the mechanical
frequency $\Omega $ in contrast to the prediction of the linear treatment. A
process of frequency up-conversion thus takes place in the opto-mechanical
coupling between vacuum fluctuations and mechanical motion of scatterers.
Second the spectrum always vanishes for all multiple integers of the
mechanical oscillation frequency. Due to the opto-mechanical resonance
condition motion-induced radiation is furthermore only predicted at
particular frequencies corresponding to fractions of the mechanical
oscillation frequency. These signatures are different from pick-up effects
and could serve to identify motion-induced radiation.

So far we have discussed the behavior of the system in a regime where the
cavity amplifies the dissipative effects of vacuum fluctuations. However, as
a consequence of the divergence of the energy density there exists a
threshold above which the system will show self-sustained oscillations in
analogy with an optical parametric oscillator\cite{OPO}. This regime is
reached if the cavity finesse is increased above its threshold value. The
amplification of motion-induced radiation should then exceed the cavity
losses. It is to be expected that in this regime the cavity will emit photon
pulses with much larger intensity than below threshold. If it were possible
to reach this regime experimentally an observation of motion-induced
radiation as well as of its characteristics could be achieved more easily.
It might thus also be interesting to calculate the radiated energy and the
spectrum above threshold. The question then arises which are the mechanisms
limiting the amplification of radiation in this regime.

In conclusion, these results confirm the idea that it might be possible to
show experimental evidence of the dissipative effects of motion in quantum
vacuum.

\noindent {\bf Acknowledgements} We thank M.~Brune, M.~Devoret, D.~Est\`eve,
S.~Haroche, J.-M.~Raimond and C.~Urbina for fruitful discussions.

\end{document}